\documentclass[twocolumn,showpacs,nofootinbib,aps,prd]{revtex4}
\usepackage{graphicx,dcolumn,bm,amsfonts,amsmath,color,xcolor}
\definecolor{blackk}{rgb}{0,0.4,0.4}
\usepackage[colorlinks=true,citecolor=blue, linkcolor=blue,urlcolor=blue]{hyperref}
\usepackage{slashed}
\usepackage{ulem}
\definecolor{bluee}{rgb}{0,0,0.7}
\newcommand\modi[1]{{\color{black}#1}}

\allowdisplaybreaks
\begin{document}

\title{$\rho$-meson longitudinal leading-twist distribution amplitude revisited and the $D\to \rho$ semileptonic decay}
\author{Tao Zhong}
\email{zhongtao1219@sina.com}
\address{Department of Physics, Guizhou Minzu University, Guiyang 550025, P.R. China}
\author{Ya-Hong Dai}
\address{Department of Physics, Guizhou Minzu University, Guiyang 550025, P.R. China}
\author{Hai-Bing Fu\footnote{Corresponding author}}
\email{fuhb@gzmu.edu.cn}
\address{Department of Physics, Guizhou Minzu University, Guiyang 550025, P.R. China}

\date{\today}
\begin{abstract}
Motivated by our previous work [Phys. Rev. D \textbf{104}, no.1, 016021 (2021)] on pionic leading-twist distribution amplitude (DA), we revisit $\rho$-meson leading-twist longitudinal DA $\phi_{2;\rho}^\|(x,\mu)$ in this paper. A model proposed by Chang based on the Dyson-Schwinger equations (DSEs) is adopted to describe the behavior of $\phi_{2;\rho}^\|(x,\mu)$. On the other hand, the $\xi$-moments of $\phi_{2;\rho}^\|(x,\mu)$ are calculated with the QCD sum rules in the framework of the background field theory. The sum rule formula for those moments are improved. More accurate values for the first five nonzero $\xi$-moments at typical scale $\mu =1, 1.4, 2, 3~{\rm GeV}$ are given, e.g., at $\mu = 1~{\rm GeV}$, \modi{$\langle\xi^2\rangle_{2;\rho}^\| = 0.220(6) $, $\langle\xi^4\rangle_{2;\rho}^\| = 0.103(4)$, $\langle\xi^6\rangle_{2;\rho}^\| = 0.066(5)$, $\langle\xi^8\rangle_{2;\rho}^\| = 0.046(4)$ and $\langle\xi^{10}\rangle_{2;\rho}^\| = 0.035(3)$}. By fitting those values with the least squares method, the DSE model for $\phi_{2;\rho}^\|(x,\mu)$ is determined. By taking the left-handed current light-cone sum rule approach, we get the transition form factor at large recoil region, {\it i.e.} $A_1(0) = 0.498^{+0.014}_{-0.012}$, $A_2(0)=0.460^{+0.055}_{-0.047}$, $V(0) = 0.800^{+0.015}_{-0.014}$, and the ratio $r_2 = 0.923^{+0.133}_{-0.119}$, $r_V = 1.607^{+0.071}_{-0.071}$. After making the extrapolation with a rapidly converging series based on $z(t)$-expansion, we present the $|V_{cd}|$-independent decay width for the semileptonic decays $D\to\rho\ell^+\nu_\ell$. Finally, the branching fractions are  $\mathcal{B}(D^0\to \rho^- e^+ \nu_e) = 1.889^{+0.176}_{-0.170}\pm 0.005$, $\mathcal{B}(D^+ \to \rho^0 e^+ \nu_e) = 2.380^{+0.221}_{-0.214}\pm 0.012$, $\mathcal{B}(D^0\to \rho^- \mu^+ \nu_\mu) = 1.881^{+0.174}_{-0.168}\pm 0.005$, $\mathcal{B}(D^+ \to \rho^0 \mu^+ \nu_\mu) =2.369^{+0.219}_{-0.211}\pm 0.011$.
\end{abstract}

\maketitle

\section{Introduction}\label{Sec:1}
The $\rho$-meson leading-twist longitudinal distribution amplitude (DA) is the key input parameter for investigating related exclusive processes including $\rho$ meson. The charmed semileptonic decay processes can provide a clear platform in researching $\rho$-meson DA. In the charmed factories, the BESIII collaboration report a newly results for semileptonic decay $D^0\to \rho^-\mu^+\nu_\mu$ in 2021~\cite{BESIII:2021pvy}. In 2019, BESIII collaboration given the improved measurements of $D\to\rho e^+\nu_e$~\cite{BESIII:2018qmf}. The CLEO collaboration provides measurement of the form factors in decays $D^{0/+}\to \rho^{-/0}e^+\nu_e$ in 2013~\cite{CLEO:2011ab} and their previous measurements~\cite{CLEO:2005cuk, CLEO:2005rxg}. It is known that, the transition form factors (TFFs) are the key component for the semileptonic $D \to \rho \ell^+ \nu_\ell$ decays in the standard model. Therefore, an accurate TFFs is very important for theoretical groups and experimental collaborations. Theoretically, the $D\to\rho$ TFFs can be treated with different approaches, such as 3-point sum rule (3PSR)~\cite{Ball:1993tp}, heavy quark effective field theory (HQEFT)~\cite{Wang:2002zba, Wu:2006rd}, relativistic harmonic oscillator potential model (RHOPM)~\cite{Wirbel:1985ji}, quark model (QM)~\cite{Isgur:1988gb, Melikhov:2000yu}, light-front quark model (LFQM)~\cite{Verma:2011yw,Cheng:2017pcq}, light-cone sum rule (LCSR)~\cite{Leng:2020fei,Fu:2018yin,Gao:2019lta}, covariant confined quark model (CCQM)~\cite{Soni:2018adu}, heavy meson and chiral symmetries theory(HM$\chi$T)~\cite{Fajfer:2005ug}, Lattice QCD (LQCD)~\cite{Lubicz:1991bi, Bernard:1991bz}. The LCSR approach is based on the operator production expansion near the light-cone, and all the non-perturbative dynamics are parameterize into the light-cone distribution amplitudes (LCDAs), which is suitable in calculating the heavy to light transition. In this paper, we will take the LCSR approach to calculate the $D\to\rho$ TFFs with left-handed current\footnote{Chiral current LCSR approach will be introduced in the next section of this paper}. Thus the key task falls to an accurate $\rho$-meson longitudinal twist-2 DA.

Theoretically, the $\rho$-meson longitudinal leading-twist DA, can be researched by several different methods, such as the QCD sum rules (SRs)~\cite{Ball:1996tb, Bakulev:1998pf, Ball:2007rt, Ball:2007zt, Pimikov:2013usa, Stefanis:2015qha},LQCD~\cite{Boyle:2008nj, Arthur:2010xf, Segovia:2013eca, Braun:2016wnx}, AdS/QCD holographic method~\cite{Forshaw:2012im, Forshaw:2012mb, Ahmady:2012dy, Ahmady:2013cga}, extracting from the experimental data~\cite{Forshaw:2010py, Forshaw:2010fz, Forshaw:2011xy, Forshaw:2011yj}, light-front quark model (LFQM)~\cite{Choi:2007yu, Dhiman:2019ddr}, Dyson-Schwinger equations (DSEs)~\cite{Gao:2014bca}, the large momentum effective theory (LMET)~\cite{Xu:2018mpf}, the instanton vacuum~\cite{Dorokhov:2006xw, Polyakov:2020cnc}, and other models~\cite{Fu:2014cna, Almeida-Zamora:2023rwg}. The QCDSR in the framework of background field theory (BFT) is an effective approach in calculating light or heavy meson DA~\cite{Huang:1989gv}. Early in 2016, we have preliminary study the $\rho$-meson longitudinal leading-twist DA~\cite{Fu:2016yzx}. In which, the first two nonzero $\xi$-moments and Gegenbauer moments are obtained, and the behavior of $\phi_{2;\rho}^\|(x,\mu)$ described by the light-cone harmonic oscillator (LCHO) model is determined~\cite{Wu:2013lga}.

Recently in 2021, we proposed a new research scheme for the QCD SR study on pionic leading-twist DAs, where a new sum rule formula for $\xi$-moments is proposed after considering the fact that the sum rule for zeroth $\xi$-moment cannot be normalized in entire Borel region~\cite{Zhong:2021epq}. Meanwhile, it enables us to calculate more higher-order $\xi$-moments. Further, the behavior of pion DA $\phi_{2;\pi}(x,\mu)$ can be determined by fitting enough $\xi$-moments with the least squares method. Subsequently, this scheme was used to study pseudoscalar $\eta^{(\prime)}$-meson and kaon leading-twist DA~\cite{Hu:2021zmy, Zhong:2022ecl} and $D$-meson twist-2,3 DAs~\cite{Zhong:2022ugk}, axial vector $a_1(1260)$-meson twist-2 longitudinal DA~\cite{Hu:2021lkl}, scalar $a_0(980)$ and $K_0^\ast(1430)$-meson leading-twist DAs~\cite{Wu:2022qqx, Huang:2022xny}. Inspired by this, we will restudy $\rho$-meson leading-twist longitudinal DA $\phi_{2;\rho}^\|(x,\mu)$ by adopting the research scheme proposed in Ref.~\cite{Zhong:2021epq} in this paper.

The remaining parts of the paper are organized as follows. In Sec.~\ref{section:2}, we present the calculation technology for $\xi$-moments of $\rho$-meson leading-twist DA, the $D\to \rho$ TFFs and the semileptonic decays $D\to\rho\ell^+\nu_\ell$. In Sec.~\ref{section:3}, we present the numerical results and discussions on $\xi$-moments, $D\to \rho$ TFFs, $D\to \rho\ell^+\nu_\ell$ decay widths and branching ratios . Section~\ref{section:4} is reserved for a summary.

\section{Theoretical framework}\label{section:2}
In order to derive the sum rules of $\xi$-moments of $\rho$-meson leading-twist longitudinal DA, we adopt the following correlation function (correlator),
\begin{align}
\Pi(z,q) &= i\int d^4x e^{iq\cdot x} \langle 0 | T\{J_n(x) J_0^\dagger(0) \} |0\rangle \nonumber\\
&= (z\cdot q)^{n+2} I(q^2), \label{eq:correlator}
\end{align}
where the current $J_n(x) = \bar{d} \slashed{z} (iz\cdot \tensor{D})^n u(x)$ with $z^2 = 0$. Following the standard calculation procedure of the QCD SRs in the framework of BFT~\cite{Zhong:2021epq}, the sum rule of $\langle\xi^n\rangle_{2;\rho}^\| \times \langle\xi^0\rangle_{2;\rho}^\|$ reads
\begin{align}
& \frac{\langle\xi^n\rangle_{2;\rho}^\| \langle\xi^0\rangle_{2;\rho}^\| f_\rho^2}{M^2 e^{m_\rho^2/M^2}} \nonumber\\
&= \frac{3}{4\modi{\pi}^2} \frac{1}{(n+1)(n+3)} (1 - e^{-s_\rho/M^2}) + \frac{(m_d + m_u) \langle\bar{q}q\rangle}{(M^2)^2} \nonumber\\
&+ \frac{\langle\alpha_s G^2\rangle}{(M^2)^2} \frac{1 + n\theta(n-2)}{12\pi(n+1)} - \frac{(m_d + m_u) \langle g_s\bar{q}\sigma TGq\rangle}{(M^2)^3} \frac{8n+1}{18} \nonumber\\
&+ \frac{\langle g_s\bar{q}q \rangle^2}{(M^2)^3} \frac{4(2n+1)}{81} - \frac{\langle g_s^3fG^3\rangle}{(M^2)^3} \frac{n\theta(n-2)}{48\pi^2} + \frac{\langle g_s^2\bar{q}q\rangle^2}{(M^2)^2} \frac{2+\kappa^2}{486\pi^2} \nonumber\\
&\times \Big\{ -2(51n+25) \Big( -\ln \frac{M^2}{\mu^2} \Big) + 3(17n+35) + \theta(n-2) \nonumber\\
&\times \Big[ 2n\Big( -\ln \frac{M^2}{\mu^2} \Big) + \frac{49n^2 + 100n + 56}{n} - 25(2n+1) \nonumber\\
&\times \Big[ \psi \Big( \frac{n+1}{2} \Big) - \psi \Big( \frac{n}{2} \Big) + \ln 4 \Big] \Big] \Big\},
\label{eq:srxinxi0}
\end{align}
where $m_\rho$ and $f_\rho$ are the $\rho$-meson mass and decay constant, $s_\rho$ is the continuum threshold, $m_u$ and $m_d$ are the current quark masses of $u$ and $d$ quark, $M$ is the Borel parameter, $\langle\bar{q}q\rangle$ with $q = u (d)$ is the double-quark condensate, $\langle\alpha_sG^2\rangle$ is the double-gluon condensate, $\langle g_s\bar{q}\sigma TGq\rangle$ is the quark-gluon mix condensate, $\langle g_s^3fG^3\rangle$ is the triple-gluon condensate, $\langle g_s\bar{q}q \rangle^2$ and $\langle g_s^2\bar{q}q \rangle^2$ are the four-quark condensates, respectively. In addition, $\kappa = \langle\bar{s}s\rangle / \langle\bar{q}q\rangle$ with the double $s$ quark condensate $\langle\bar{s}s\rangle$. By taking $n = 0$ for Eq.~\eqref{eq:srxinxi0}, the sum rule of zeroth $\xi$-moment $\langle\xi^0\rangle_{2;\rho}^\|$ can be obtained,
\begin{align}
\frac{\modi{(\langle\xi^0\rangle_{2;\rho}^\|)^2} f_\rho^2}{M^2 e^{m_\rho^2/M^2}} &= \frac{1}{4\pi^2} (1 - e^{-s_\rho/M^2}) + (m_d + m_u) \frac{\langle\bar{q}q\rangle}{(M^2)^2} \nonumber\\
&+ \frac{\langle\alpha_s G^2\rangle}{(M^2)^2} \frac{1}{12\pi} - \frac{1}{18} (m_d + m_u) \frac{\langle g_s\bar{q}\sigma TGq\rangle}{(M^2)^3} \nonumber\\
&+ \frac{4}{81} \frac{\langle g_s\bar{q}q \rangle^2}{(M^2)^3} + \frac{\langle g_s^2\bar{q}q\rangle^2}{(M^2)^2} \frac{2+\kappa^2}{486\pi^2} \nonumber\\
&\times \Big[ -50 \Big( -\ln \frac{M^2}{\mu^2} \Big) + 105 \Big].
\label{eq:srxi0xi0}
\end{align}
The Eq.~\eqref{eq:srxi0xi0} indicates that the zeroth $\xi$-moment $\langle\xi^0\rangle_{2;\rho}^{\|}$ in Eq.~\eqref{eq:srxinxi0} cannot be normalized in the entire Borel parameter region due to that the contributions of vacuum condensates with dimension greater than six are truncated. As discussed in Ref.~\cite{Zhong:2021epq}, more accurate and reasonable sum rule for $n$th $\xi$-moment $\langle\xi^n\rangle_{2;\rho}^\|$ should be
\begin{align}
\langle\xi^n\rangle_{2;\rho}^\| = \frac{\langle\xi^n\rangle_{2;\rho}^\| \times \langle\xi^0\rangle_{2;\rho}^\| |_{\rm From\ Eq.~\eqref{eq:srxinxi0}}}{\sqrt{(\langle\xi^0\rangle_{2;\rho}^\|)^2} |_{\rm From\ Eq.~\eqref{eq:srxi0xi0}}}
\label{eq:srxin}
\end{align}

On the other hand, to describe the behavior of $\rho$-meson leading-twist longitudinal DA, we take the following DSE model for $\phi_{2;\rho}^\parallel(x,\mu)$~\cite{Zhong:2022lmn, Chang:2013pq},
\begin{align}
\phi_{2;\rho}^\parallel(x,\mu) = \mathcal{N} [x(1-x)]^{\alpha_-} \Big[ 1 + \hat{a}_2 C_2^\alpha(2x-1) \Big],
\label{eq:DA_DSE}
\end{align}
where $\alpha_- = \alpha - 1/2$, and $\mathcal{N} = 4^\alpha \Gamma(\alpha + 1) / [\sqrt{\pi} \Gamma(\alpha + 1/2)]$ is the normalization constant.

Nextly, in order to get the $D\to\rho$ TFFs, one can take the following correlation function
\begin{align}
&\Pi_\mu(p,q) = i\int d^4x e^{iq\cdot x}   \nonumber\\
&\times \langle\rho (\tilde p,\tilde\epsilon)|{\rm T} \big\{\bar q_1(x)\gamma_\mu(1-\gamma_5)c(x),  i \bar c(0)(1 - \gamma_5)q_2(0)\big\} |0\rangle, \nonumber\\ \label{correlator}
\end{align}
with the $j_D^{L} (x)=i \bar q_2(x)(1 - \gamma_5)c(x)$ is the left-handed current. As we know, there will be fifteen DAs for vector meson up to twist-4 accuracy, the left-handed chiral current can reduce the uncertainties from chiral-odd vector meson DAs with $\delta^{0,2}$-order and leave the chiral-even with $\delta^{1,3}$-order meson. The relationship is also listed in Table~\ref{DA_delta}. In this table, except $j_D^{L} (x)$, the current $j_D^{R} (x)$ respect right-handed current with expression $j_D^{R} (x)=i \bar q_2(x)(1 + \gamma_5)c(x)$, which have been researched in our previous work~\cite{Fu:2018yin}. The parameter $\delta \simeq m_\rho/m_c\sim 52\%$~\cite{Ball:2004rg, Ball:1998sk}. Meanwhile, the chiral current approach have been considered in some references~\cite{Huang:1998gp, Huang:2001xb, Wan:2002hz, Zuo:2006dk, Wu:2007vi, Wu:2009kq}, which improve the predictions of the LCSR approach.
\begin{table}[t]
\footnotesize
\centering
\caption{The $\rho$-meson DAs with different twist-structures up to $\delta^3$, where $\delta \simeq m_\rho/m_c$.} \label{DA_delta}
\begin{tabular}{|l| c |c |c|c|}
\hline
& &twist-2 &twist-3 & twist-4  \\
\hline
                                                             & $\delta^0$      & $\phi_{2;\rho}^\bot$  &  &  \\
\raisebox {1.5ex}[0pt]{$j_D^{L} (x)$}  &$\delta^2$         &  & $\phi_{3;\rho}^\|, \psi_{3;\rho}^\|, \Phi_{3;\rho}^\bot$ & $  \phi_{4;\rho}^\bot ,\psi_{4;\rho}^\bot,\Psi_{4;\rho}^\bot, \widetilde{\Psi} _{4;\rho}^\bot$\\\hline
                                                             & $\delta^1$         & $\phi_{2;\rho}^\|$ & $\phi_{3;\rho}^\bot, \psi_{3;\rho}^\bot, \Phi_{3;\rho}^\|,\tilde\Phi_{3;\rho }^\bot$  &   \\
\raisebox {1.5ex}[0pt]{$j_D^{R} (x)$}  &$\delta^3$         &  &  & $\phi_{4;\rho}^\|,\psi_{4;\rho}^\|$\\
\hline
\end{tabular}
\end{table}

\begin{table*}
\footnotesize
\caption{Our predictions for the first five nonzero $\xi$-moments $\langle \xi^n\rangle _{2;\rho}^\parallel(n=2,\cdots,10)$ and the second Gegenbauer moment $a_{2 \parallel}^{2;\rho}$ of the $\rho$-meson leading-twist longitudinal DA, compared to other theoretical predictions.}\label{t:xin_value}
\begin{tabular}{l c c c c c c c}
\hline\hline
& ~$\mu{\rm [GeV]}$~ & ~$~~~~~\langle\xi^2\rangle_{2;\rho}^\parallel~~~~~$~ & ~$~~~~~\langle\xi^4\rangle_{2;\rho}^\parallel~~~~~$~ & ~$~~~~~\langle\xi^6\rangle_{2;\rho}^\parallel~~~~~$~ & ~$~~~~~\langle\xi^8\rangle_{2;\rho}^\parallel~~~~~$~ & ~$~~~~~\langle\xi^{10}\rangle_{2;\rho}^\parallel~~~~~$~ & ~$~~~~~a_{2 \parallel}^{2;\rho}~~~~~$  \\ \hline
This Work               & 1  & $0.220(6)$ & $0.103(4)$ & $0.0656(50)$ & $0.0457(35)$ & $0.0346(28)$ & $0.059(18)$ \\
This Work               & 1.4  & $0.217(5)$ & $0.100(3)$ & $0.0623(40)$ & $0.0428(29)$ & $0.0320(23)$ & $0.051(16)$ \\
This Work               & 2  & $0.215(5)$ & $0.0986(31)$ & $0.0603(35)$ & $0.0411(25)$ & $0.0304(20)$ & $0.045(14)$ \\
This Work               & 3  & $0.214(4)$ & $0.0972(28)$ & $0.0587(30)$ & $0.0397(22)$ & $0.0292(18)$ & $0.041(13)$ \\
QCD SRs~\cite{Ball:1996tb}               & 1  & $$ & $$ & $$ & $$ & $$ & $0.18(10)$ \\
QCD SRs~\cite{Bakulev:1998pf}               & 1  & $0.227(7)$ & $0.095(5)$ & $0.051(4)$ & $0.030(2)$ & $0.020(5)$ & $$ \\
QCD SRs~\cite{Ball:2007rt, Ball:2007zt}               & 1  & $$ & $$ & $$ & $$ & $$ & $0.15(7)$ \\
QCD SRs~\cite{Ball:2007rt, Ball:2007zt}               & 2  & $$ & $$ & $$ & $$ & $$ & $0.10(5)$ \\
QCD SRs~\cite{Pimikov:2013usa}               & 1  & $0.216(21)$ & $0.089(9)$ & $0.048(5)$ & $0.030(3)$ & $0.022(2)$ & $0.047(58)$ \\
QCD SRs~\cite{Stefanis:2015qha}               & 2  & $0.206(8)$ & $0.087(6)$ & $$ & $$ & $$ & $0.017(24)$ \\
LQCD~\cite{Boyle:2008nj}               & 2  & $0.237(36)(12)$ & $$ & $$ & $$ & $$ & $$ \\
LQCD~\cite{Arthur:2010xf, Segovia:2013eca}               & 2  & $0.27(1)(2)$ & $$ & $$ & $$ & $$ & $$ \\
LQCD~\cite{Braun:2016wnx}               & 2  & $$ & $$ & $$ & $$ & $$ & $0.132(27)$ \\
AdS/QCD~\cite{Forshaw:2012im, Forshaw:2012mb}               & 1  & $0.228$ & $$ & $$ & $$ & $$ & $$ \\
Data Fitting~\cite{Forshaw:2010py, Forshaw:2010fz}               & 1  & $0.227$ & $0.105$ & $0.062$ & $0.041$ & $0.029$ & $$ \\
LFQM~\cite{Choi:2007yu}               & 1  & $0.21,0.19$ & $0.09,0.08$ & $0.05,0.04$ & $$ & $$ & $0.02,-0.02$ \\
DSE~\cite{Gao:2014bca}               & 2  & $0.23$ & $0.11$ & $0.066$ & $0.045$ & $0.033$ & $$ \\
\hline\hline
\end{tabular}
\end{table*}
Based on the procedures of LCSR approach, one will get the $D\to \rho$ TFFs $A_{1,2}(q^2)$ and $V(q^2)$ LCSRs expression with the left-handed current in the correlator. The analytic formulas are similar with the $B\to\rho$ TFFs of our previous work~\cite{Fu:2014cna}, which make a replacement for the input parameter of $B$-meson with $D$-meson such as $m_B\to m_D$, $f_B \to f_D$, $m_b \to m_c$. Thus, we do not listed here in this paper. Moreover, there have two ratio $r_V = {V(0)}/{A_1(0)}$ and $r_2={A_2(0)}/{A_1(0)}$ based on the TFFs $A_{1,2}(q^2)$ and $V(q^2)$, which have less uncertainties between the different approaches.

Furthermore, the semileptonic decay width for $D\to \rho \ell^+ \nu_\ell$ are composed by three different parts, which can be expressed as,
\begin{align}
\Gamma &=\frac{G_F^2 C_k|V_{cd}|^2 }{192\pi^3 m_D^3}\int_{m_\ell^2}^{q^2_{\rm max}}q^2\sqrt{\lambda(q^2)}
\nonumber\\
&\times \Big\{ |H_+(q^2)|^2+|H_-(q^2)|^2+|H_0(q^2)|^2\Big\}, \label{Eq:GammaLT0}
\end{align}
where the constant $C_k = 1$ for the $D^0\to\rho^-\ell^+\nu_\ell$ and $C_k = 1/\sqrt{2}$ for $D^+\to\rho^0\ell^+\nu_\ell$. Other parameters and expressions have the definitions: $G_F=1.166\times10^{-5}~{\rm GeV}^{-2}$ is the Fermi constant, $\lambda(q^2) = (m_D^2 + m_\rho^2 - q^2)^2-4 m_D^2 m_\rho^2$ is the phase-space factor, $q^2_{\rm max} = (m_D-m_\rho)^2$ is the small recoil point of the $D\to\rho$ transition. The three helicity decay amplitudes $H_{\pm}(q^2)$ and $H_0(q^2)$ are mainly separated by the transition amplitude with definite spin-parity quantum number, which can be found in our previous work~\cite{Fu:2018yin}. Meanwhile, the longitudinal and transverse helicity amplitudes are expressed as $\Gamma^{\rm T} = \Gamma^+ + \Gamma^-$, $\Gamma^{\rm L}=\Gamma^0$ and $\Gamma=\Gamma^{\rm L} + \Gamma^{\rm T}$.
\begin{figure}[t]
\centering
\includegraphics[width=0.42\textwidth]{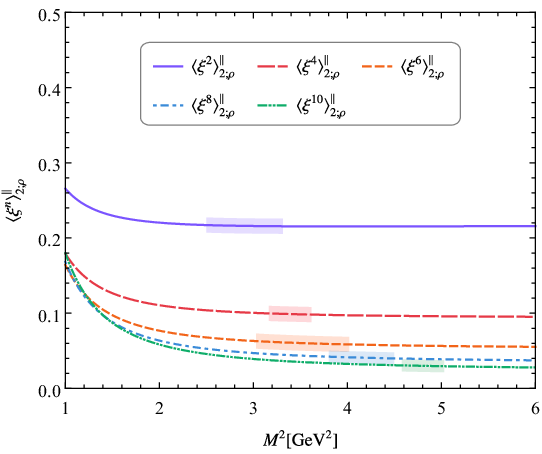}
\caption{The $\rho$-meson leading-twist longitudinal DA moments $\langle \xi^n\rangle_{2;\rho}^\parallel$ with $n=(2,\cdots,10)$ versus Borel parameter $M^2$ and the corresponding Borel windows, where all input parameters are set to be their central values.}
\label{f:xinM2}
\end{figure}
\section{Numerical analysis}\label{section:3}

To perform the numerical calculation, we take $f_\rho = 210 \pm 4~{\rm MeV}$~\cite{Dhiman:2019ddr, ParticleDataGroup:2018ovx}, $m_\rho = 775.26 \pm 0.23~{\rm MeV}$, the current quark masses $m_u = 2.16^{+0.49}_{-0.26}~{\rm MeV}$ and $m_d = 4.67^{+0.48}_{-0.17}~{\rm MeV}$~\cite{Workman:2022ynf}, the vacuum condensates $\langle\bar{q}q\rangle = (-2.417^{+0.227}_{-0.114}) \times 10^{-2}~{\rm GeV}^2$, $\langle g_s\bar{q}\sigma TGq\rangle = (-1.934^{+0.188}_{-0.103}) \times 10^{-2}~{\rm GeV}^5$, $\langle g_s\bar{q}q\rangle^2 = (2.082^{+0.734}_{-0.697}) \times 10^{-3}~{\rm GeV}^6$, $\langle g_s^2\bar{q}q\rangle^2 = (7.420^{+2.614}_{-2.483}) \times 10^{-3}~{\rm GeV}^6$, $\langle\alpha_s G^2\rangle = 0.038 \pm 0.11~{\rm GeV}^4$, $\langle g_s^3fG\rangle \simeq 0.045~{\rm GeV}^6$ and $\kappa = 0.74 \pm 0.03$~\cite{Zhong:2021epq, Colangelo:2000dp, Narison:2014wqa}. In which, the current quark masses and the vacuum condensates other than gluon condensates are scale dependence, and whose above values are at $\mu = 2~{\rm GeV}$. For the scale evolution of those inputs, one can refer to Ref.~\cite{Zhong:2021epq}. In calculation, we take the scale $\mu = M$ as usual. By requiring that there is a reasonable Borel window to normalize $\langle\xi^0\rangle_{2;\rho}^\parallel$ with Eq.~\eqref{eq:srxi0xi0}, we obtain the continuum threshold as $s_\rho \simeq 2.1~{\rm GeV}^2$.

Now, one can obtain the $\xi$-moment versus Borel parameter. Further, we can determine the Borel window and then the value of $n$th $\xi$-moment $\langle\xi^n\rangle_{2;\rho}^\parallel$ based on the well-known criteria that the continuum state's contribution and dimension-six condensate's contribution are as small as possible, and the value of $\langle\xi^n\rangle_{2;\rho}^\parallel$ is stable in the Borel window. Specifically, the continuum contribution are not more than $30\%$ for all $n$th-order; the dimension-six contribution of $\langle\xi^n\rangle_{2;\rho}^\parallel$ are less than $2\%$ for $n = (2,4)$, $5\%$ for $n = (6,8,10)$, respectively. It should be noted that, only the even order $\xi$-moments are not zero due to the isospin symmetry. The $\xi$-moments versus Borel parameter and the obtained Borel windows are shown in Fig.~\ref{f:xinM2}, where the shaded areas are the Borel windows for $n=(2,\cdots,10)$ respectively.

Then, the values of ${\langle\xi^n\rangle_{2;\rho}^\parallel}|_\mu$ up to 10th order with four different scales $\mu = 1,1.4,2,3{\rm GeV}$ and the second Gegenbauer moment $a_{2;\rho}^\parallel$ can be obtained and exhibited in Table~\ref{t:xin_value}. Here, we only calculate the values of the first five non-zero $\xi$-moments in this work because, as shown in Ref.~\cite{Zhong:2022lmn}, these moments are sufficient to determine the behavior of DA $\phi_{2;\rho}^\parallel(x,\mu)$. As a comparison, the other theoretical predictions obtained by QCD SRs~\cite{Ball:1996tb, Bakulev:1998pf, Ball:2007rt, Ball:2007zt, Pimikov:2013usa, Stefanis:2015qha}, LQCD~\cite{Boyle:2008nj, Arthur:2010xf, Segovia:2013eca, Braun:2016wnx}, AdS/QCD~\cite{Forshaw:2012im, Forshaw:2012mb}, Data Fitting~\cite{Forshaw:2010py, Forshaw:2010fz}, LFQM~\cite{Choi:2007yu} and DSE~\cite{Gao:2014bca} are also shown in Table~\ref{t:xin_value}. One can find that, our second moment is less than the QCD SR predictions in Ref.~\cite{Ball:1996tb, Ball:2007rt, Ball:2007zt}, LQCD calculations~\cite{Boyle:2008nj, Arthur:2010xf, Segovia:2013eca, Braun:2016wnx} and DSE result~\cite{Gao:2014bca}, and larger than QCD SR predictions in Ref.~\cite{Pimikov:2013usa, Stefanis:2015qha}, but very consistant with the value with QCD SRs in Ref.~\cite{Bakulev:1998pf}, the AdS/QCD prediction~\cite{Forshaw:2012im, Forshaw:2012mb} and the value extracted from the HERA data on diffractive $\rho$ photoproduction~\cite{Forshaw:2010py, Forshaw:2010fz}.

\begin{table}[t]
\footnotesize
\caption{The model parameters $\alpha$ and $\hat{a}_2$ of our DA obtained by fitting using the least squares method and the corresponding $\chi^2_{\rm min}/n_d$ and goodness of fit $P_{\chi^2_{\rm min}}$ at several typical scale such as $\mu = 1, 1.4, 2, 3~{\rm GeV}$.} \label{t:ModelParameter}
\begin{tabular}{ c c c c c }
\hline\hline
~~~~$\mu$~~~~~ & ~~~~~$\alpha$~~~~~ & ~~~~~$\hat{a}_2$~~~~~ & ~~~~~$\chi^2_{\rm min}/n_d$~~~~~ & ~~~~~$P_{\chi^2_{\rm min}}$~~~~ \\
\hline
~$1$~ & ~$0.625$~ & ~$-0.516$~ & ~$0.957/3$~ & ~$0.812$~ \\
~$1.4$~ & ~$0.678$~ & ~$-0.448$~ & ~$0.488/3$~ & ~$0.922$~ \\
~$2$~ & ~$0.721$~ & ~$-0.398$~ & ~$0.259/3$~ & ~$0.968$~ \\
~$3$~ & ~$0.709$~ & ~$-0.427$~ & ~$0.141/3$~ & ~$0.987$~ \\
\hline\hline
\end{tabular}
\end{table}

\begin{figure}[t]
\centering
\includegraphics[width=0.42\textwidth]{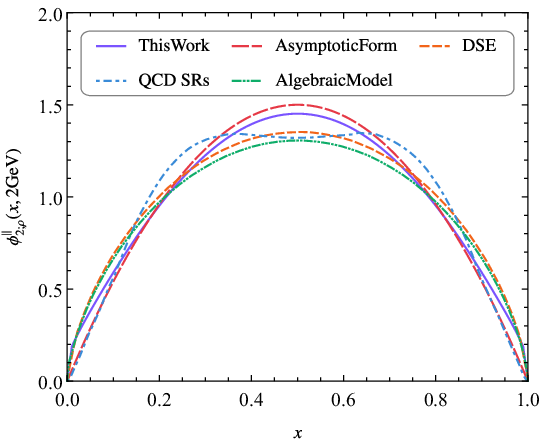}
\caption{Curve of our $\rho$-meson leading-twist longitudinal DA. As a comparison, the DA curves obtained by DSE~\cite{Gao:2014bca}, QCD SRs~\cite{Stefanis:2015qha}, algebraic model~\cite{Almeida-Zamora:2023rwg} and the asymptotic form are also shown.}
\label{f:DA}
\end{figure}

By fitting our values of $\langle\xi^n\rangle_{2;\rho}^\parallel (n = 2,4,6,8,10)$ shown in Table~\ref{t:xin_value} with the least squares method (where the model parameters $\alpha$ and $\hat{a}_2$ are taken to be the fitting parameters, and for specific fitting procedure, one can refer to Ref.~\cite{Zhong:2021epq}), the behavior of the $\rho$-meson leading-twist longitudinal DA can be obtained. Specifically, the model parameters $\alpha$ and $\hat{a}_2$ of our DA and the corresponding $\chi^2_{\rm min}/n_d$ and goodness of fit $P_{\chi^2_{\rm min}}$ at several typical scale such as $\mu = 1, 1.4, 2, 3~{\rm GeV}$ are exhibited in Table~\ref{t:ModelParameter}. Then, the curve of $\rho$-meson leading-twist longitudinal DA is shown in Fig.~\ref{f:DA}. Meanwhile, other predictions in literature~\cite{Stefanis:2015qha, Gao:2014bca, Almeida-Zamora:2023rwg} are also shown for comparison. One can find that our curve is closer to the asymptotic form.

\begin{table}[b]
\footnotesize
\centering
\caption{The $D\to \rho$ TFFs $A_{1,2}(q^2)$ and $V(q^2)$ at the large recoil region $q^2\simeq 0$. The errors are squared averages of all the mentioned error sources. As a comparison, we also present the prediction from various methods.}
\label{T:TFF0}
\begin{tabular}{llll}
\hline\hline
& $A_1(0)$ & $A_2(0)$ & $V(0)$ \\
\hline
    This work & $0.498^{+0.014}_{-0.012}$ &	$0.460^{+0.055}_{-0.047}$ & $0.800^{+0.015}_{-0.014}$  \\
    CLEO2013 \cite{CLEO:2011ab} & $0.56(1)^{+0.02}_{-0.03}$ &	$0.47(6)(4)$ & $0.84(9)^{+0.05}_{-0.06}$  \\
    3PSR~\cite{Ball:1993tp}	& $0.5(2)$	& $0.4(1)$ & 	$1.0(2)$ \\
    HQETF~\cite{Wang:2002zba} & $0.57(8)$ & $0.52(7)$ & $0.72(10)$\\
    LCSR~\cite{Wu:2006rd} & $0.599^{+0.035}_{-0.030}$ & $0.372^{+0.026}_{-0.031}$ & $0.801^{+0.044}_{-0.036}$\\
    RHOPM \cite{Wirbel:1985ji}& 	0.78 & 	0.92	& 1.23 \\
    QM-I \cite{Isgur:1988gb}& 	0.59& 	0.23	& 1.34  \\
    QM-II~\cite{Melikhov:2000yu} & 0.59 & 0.49 & 0.90 \\
    LFQM~\cite{Verma:2011yw}       &  $0.60(1)$   &  $0.47(0)$  &  $0.88(3)$ \\
    HM$\chi$T~\cite{Fajfer:2005ug} & 0.61 & 0.31 & 1.05 \\
    LQCD \cite{Lubicz:1991bi} & 	$0.45(4)$ & $0.02(26)$ & $0.78(12)$  \\
    LQCD \cite{Bernard:1991bz}& 	$0.65(15){^{+0.24}_{-0.23}}$ & $0.59(31){^{+0.28}_{-0.25}}$ & 	 $1.07(49)(35)$ \\
\hline\hline
\end{tabular}
\end{table}
\begin{figure*}[t]
\centering
\includegraphics[width=0.45\textwidth]{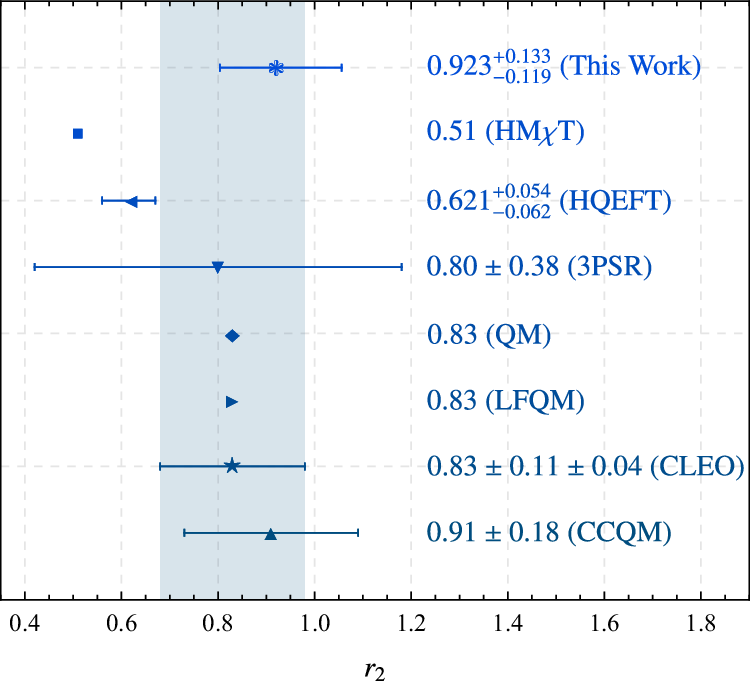}~\includegraphics[width=0.45\textwidth]{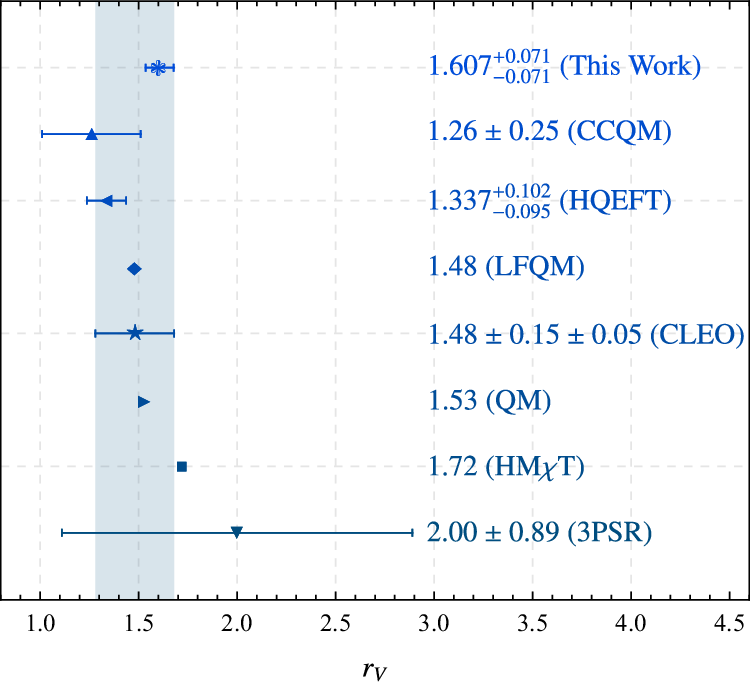}
\caption{Predictions for the ratios $r_2$ and $r_V$ within uncertainties. The CLEO collaboration~\cite{CLEO:2011ab}, HM$\chi$T~\cite{Fajfer:2005ug}, LFQM~\cite{Verma:2011yw}, HQEFT~\cite{Wang:2002zba},  3PSR~\cite{Ball:1993tp}, QM~\cite{Isgur:1988gb, Melikhov:2000yu}, CCQM~\cite{Soni:2018adu} are presented as comparison.} \label{F:TFFsratio}
\end{figure*}

Furthermore, in order to calculated the $D\to \rho$ TFFs, the input parameters should be clarified. The current charm-quark mass is $m_c = 1.27\pm0.02~{\rm GeV}$, $D$-meson mass is $m_D = 1.869~{\rm GeV}$~\cite{Workman:2022ynf}. Based on the three criteria for determining $s_0$ and $M^2$ in the LCSR approach that can be found in our previous work~\cite{Hu:2021zmy}, we have the $s_0^{A_1} = 7.0\pm0.1~{\rm GeV}^2$, $s_0^{A_2}= 6.0\pm0.1~{\rm GeV}^2$, $s_0^{V} = 7.0\pm0.1~{\rm GeV}^2$ and $M^2_{A_1} = 1.45\pm0.05~{\rm GeV}^2$, $M^2_{A_2} = 1.20\pm0.05~{\rm GeV}^2$, $M^2_{V} = 2.15\pm0.05~{\rm GeV}^2$. Then, the TFFs at the large recoil region, {\it i.e.} $A_{1,2}(0)$ and $V(0)$ are present in Table~\ref{T:TFF0}. The uncertainties are coming from the squared average for all the input parameters. In Table~\ref{T:TFF0}, the predictions from theoretical group and experimental collaborations, {\it i.e.} the CLEO collaboration~\cite{CLEO:2011ab}, 3PSR~\cite{Ball:1993tp}, HQEFT~\cite{Wang:2002zba, Wu:2006rd}, RHOPM~\cite{Wirbel:1985ji}, QM~\cite{Isgur:1988gb, Melikhov:2000yu}, LFQM~\cite{Verma:2011yw},  HM$\chi$T~\cite{Fajfer:2005ug} and Lattice QCD predictions~\cite{Lubicz:1991bi, Bernard:1991bz} are presented, respectively. The comparison about the every results from Table~\ref{T:TFF0} indicate that the TFFs of our prediction is consistent with many approaches within errors.

Then, the ratio between different $D\to\rho$ TFFs, {\it i.e.} $r_2$ and $r_V$ are present in Fig.~\ref{F:TFFsratio}. From which, the CLEO collaboration~\cite{CLEO:2011ab}, HM$\chi$T~\cite{Fajfer:2005ug}, LFQM~\cite{Verma:2011yw}, HQEFT~\cite{Wang:2002zba},  3PSR~\cite{Ball:1993tp}, QM~\cite{Isgur:1988gb, Melikhov:2000yu}, CCQM~\cite{Soni:2018adu} are presented as comparison. From the figure we can get the conclusion
\begin{itemize}
\item Our predictions within uncertainties are in the region of the CLEO collaboration results.
\item The central value with respect uncertainties, our results for $r_2$ have agreement with QM, LFQM, 3PSR, CCQM for theoretical predictions. $r_V$ have agreement with QM and 3PSR predictions.
\item The uncertainties of our predictions are $(12.9\% \sim 14.4\%)$ for $r_2$ and $4.4\%$ for $r_V$, which shows the left-handed chiral current LCSR approach can get a reasonable results.
\end{itemize}

\begin{table}[b]
\footnotesize
\begin{center}
\caption{$|V_{cd}|$-independent total decay width $1/|V_{\rm cd}|^2 \times \Gamma$, ratio for longitudinal/transverse and positive/negative helicity decay width. As comparsion, other theoretical group predictions are also given.}
\label{T:Gamma}
\begin{tabular}{llll}
\hline\hline
& $1/|V_{\rm cd}|^2 \times \Gamma$ & $\Gamma^{\rm L}/ \Gamma^{\rm T}$ & $\Gamma^+/ \Gamma^-$ \\
\hline
This paper	               & $59.90^{+5.58}_{-5.38}$ & $1.03^{+0.10}_{-0.11}$ & $0.15^{+0.02}_{-0.02}$       \\
3PSR~\cite{Ball:1993tp}      & $15.80\pm4.61$           & $1.31\pm0.11$          & $0.24\pm0.03$ \\
HQEFT~\cite{Wang:2002zba} & $71\pm14$           & $1.17\pm0.09$          & $0.29\pm0.13$ \\
RHOPM\cite{Wirbel:1985ji}  & $90.83$                  & 0.91                   & 0.19    \\
QM~\cite{Isgur:1988gb}        & $88.86$                  & 1.33                   & 0.11    \\
LQCD\cite{Lubicz:1991bi}       & $54.63\pm12.51$          & $1.86\pm0.56$          & 0.16    \\
LQCD\cite{Bernard:1991bz}      & 71.75                    & 1.10                   & 0.18    \\
\hline\hline
\end{tabular}
\end{center}
\end{table}

\begin{table*}[t]
\footnotesize
\caption{The branching ratios of the semileptonic decays $D^0\to \rho^- \ell^+ \nu_\ell$ and $D^+ \to \rho^0 \ell^+ \nu_\ell$ (in unit: $10^{-3}$). As a comparison, we also present the results from experimental collaboration and theoretical groups.}
\label{T:BF}
\begin{tabular}{l l l l l}
\hline\hline
Decay Mode~~~~~~~~~~~~~~~~~~~~	& $D^0\to \rho^- e^+ \nu_e$~~~~~~~~~~~~~          & $D^+ \to \rho^0 e^+ \nu_e$~~~~~~~~~~~~~    & $D^0\to \rho^- \mu^+ \nu_\mu$~~~~~~~~~~~~~          & $D^+ \to \rho^0 \mu^+ \nu_\mu$        \\ \hline
This paper	& $1.889^{+0.176}_{-0.170}\pm 0.005$ & $2.380^{+0.221}_{-0.214}\pm 0.012$ & $1.881^{+0.174}_{-0.168}\pm 0.005$ & $2.369^{+0.219}_{-0.211}\pm 0.011$  \\
BESIII~\cite{BESIII:2021pvy} & - & - & $1.35\pm0.09\pm0.09$ & - \\
BESIII~\cite{BESIII:2018qmf} & $1.445\pm0.058\pm0.039$ & $1.860\pm0.070\pm0.061$ & - & - \\
CLEO2013~\cite{CLEO:2011ab}     & $1.77\pm0.12\pm0.10$      	     & $2.17\pm0.12^{+0.12}_{-0.22}$   & -& -   \\
CLEO2005~\cite{CLEO:2005rxg} & $1.94\pm0.39\pm0.13$ & $2.1\pm0.4\pm0.1$ & -& - \\
3PSR \cite{Ball:1993tp}	& $0.5\pm0.1$	& - & - & -    \\
HQEFT~\cite{Wang:2002zba} & $1.4\pm0.3$ & -& -& - \\
LCSR~\cite{Wu:2006rd} & $1.81^{+0.18}_{-0.13}$ & $2.29^{+0.23}_{-0.16}$ & $1.73^{+0.17}_{-0.13}$ & $2.20^{+0.21}_{-0.16}$ \\
NWA~\cite{Shi:2017pgh}+HQEFT~\cite{Wu:2006rd}  & $1.67\pm0.27$ & $2.16\pm0.36$ & - & - \\
NWA~\cite{Shi:2017pgh}+LFQM~\cite{Verma:2011yw}   & $1.73\pm0.07$  & $2.24\pm0.09$ & - & -\\
LFQM~\cite{Cheng:2017pcq} & - & - & $1.7\pm0.2$ & - \\
LCSR~\cite{Leng:2020fei} & $1.74\pm0.25$ & $2.25\pm0.32$ & $1.65\pm0.23$ & $2.14\pm0.30$ \\
CCQM~\cite{Soni:2018adu} & 1.62 & 2.09 & 1.55 & 2.01 \\
HM$\chi$T\cite{Fajfer:2005ug} & 2.0	& 2.5 & -& -\\
$\chi$UA~\cite{Sekihara:2015iha} & 1.97 & 2.54 & 1.84 & 2.37  \\
RQM~\cite{Faustov:2019mqr} & 1.96 & 2.49 & 1.88 & 2.39 \\
ISGW2 \cite{Scora:1995ty} & 1.0 & 1.3  & -& - \\
\hline\hline
\end{tabular}
\end{table*}

With the resultant TFFs at large recoil region, their behavior at the whole physical region, {\it i.e.} $q^2_{\rm phys.} \in [0, 1.18]~\text{GeV}^2$ should be get. It is known that the LCSRs predictions are suitable in low and intermediate momentum transfer. So a rapidly converging series based on $z(t)$-expansion are used to make extrapolation in this paper~\cite{Khodjamirian:2010vf, Straub:2015ica},
\begin{eqnarray}
F_i(q^2) = \frac1{1-q^2/m_{R,i}^2} \sum_{k=0,1,2}a_k^i [z(q^2)-z(0)]^k.
\end{eqnarray}
The formula of $z(t)$ can be found in Ref.~\cite{Khodjamirian:2010vf}. $F_i$ respect three TFFs $A_{1,2}$ and $V$, respectively. The appropriate resonance masses are $m_{R,i}= 2.007~ {\rm GeV}$ for TFF $V(q^2)$ with $J^P = 1^-$ and $m_{R,i}= 2.427~{\rm GeV}$ for TFFs $A_{1,2}(q^2)$ with $J^P = 1^+$.  The parameters $a_k^i$ can be fixed by requiring $\Delta < 0.1\%$, with parameter $\Delta$ been introduced to measure the quality of extrapolation, $\Delta={\sum_t|F_i(t)-F_i^{\rm fit}(t)|} / {\sum_t|F_i(t)|}\times 100$,
where $t\in[0,1/40,\cdots,40/40]\times 0.8 {\rm GeV}^2$.

After making the extrapolated TFFs of transition $D\to\rho$, we get the CKM-independent total decay width $1/|V_{\rm cd}|^2 \times \Gamma$. Combing with the ratio for longitudinal/transverse and positive/negative helicity decay width, we present the predictions in Table~\ref{T:Gamma}. The 3PSR~\cite{Ball:1993tp},
HQEFT~\cite{Wang:2002zba}, RHOPM\cite{Wirbel:1985ji}, QM~\cite{Isgur:1988gb}, LQCD\cite{Lubicz:1991bi} and LQCD\cite{Bernard:1991bz} results are given to make a comparison.  As can be seen from the Table~\ref{T:Gamma}, there exist large gap between different theoretical group for total decay width, and our results are agree with the Lattice results within uncertainties. The longitudinal/transever helicity decay width for our prediction is agree with HQEFT, LQCD within uncertainties, which is larger than 1. The positive/negative helicity decay width for our prediction is agree with QM and LQCD within uncertainties.

Lastly, the branching fractions for the two types of $D\to\rho\ell^+\nu_\ell$ semileptonic decays are calculated.  The first one is the $D^0$-type including $D^0 \to \rho^- e^+\nu_e$ and $D^0 \to \rho^- \mu^+\nu_\mu$ decays, with which the lifetime $\tau(D^0)= 0.410\pm 0.001$ ps should be used. The second one is the $D^+$-type including $D^+ \to \rho^0 e^+\nu_e$ and $D^+ \to \rho^0 \mu^+\nu_\mu$ associated with $\tau(D^+)= 1.033\pm 0.005$ ps~\cite{Workman:2022ynf}. The CKM matrix element $|V_{cd}| = 0.225\pm0.001$~\cite{ParticleDataGroup:2018ovx}. We present the predictions in Table~\ref{T:BF}, with two uncertainties are mainly coming from the squared average of theoretical input parameters and the experimental uncertainties from the $D$-meson lifetime. To make a comparison, theoretical results for 3PSR \cite{Ball:1993tp}, HQEFT~\cite{Wang:2002zba}, Narrow Width Approximation (NWA)~\cite{Shi:2017pgh} with HQEFT~\cite{Wu:2006rd}  and LFQM~\cite{Verma:2011yw} approaches, LCSR~\cite{Wu:2006rd,Leng:2020fei}, LFQM~\cite{Cheng:2017pcq}, CCQM~\cite{Soni:2018adu}, Chiral Unitarity Approach ($\chi$UA)~\cite{Sekihara:2015iha}, RQM~\cite{Faustov:2019mqr}, HM$\chi$T~\cite{Fajfer:2005ug}, ISGW2~\cite{Scora:1995ty} and experimental results for BESIII collaboration~\cite{BESIII:2021pvy, BESIII:2018qmf}, CLEO collaboration prediction in 2013~\cite{CLEO:2011ab} and 2005~\cite{CLEO:2005rxg} are also given. The results shows that

\begin{itemize}
  \item The results obtained by our left-handed chiral LCSR approach are consistent with those obtained by other LCSR within uncertainties.
  \item Our current results are consistent with those of the CLEO collaboration, but are larger than the BESIII collaboration predictions.
  \item Compared with other theoretical groups, our results are consistent with NWA, LFQM, HQEFT, $\chi$UA, RQM, HM$\chi$T, which still have a smaller gap with 3PSR, HQEFT, $\chi$UA, ISGW2 results.
\end{itemize}

\section{Summary}\label{section:4}
In the framework of BFT with QCD SR approach, we calculated the $\xi$-moments of $\rho$-meson leading twist longitudinal DA. Based on the zeroth $\xi$-moment can not be normalized in the entire Borel region, the new sum rule formula for the $\xi$-moment is given in Eq.~\eqref{eq:srxinxi0}. The results up to tenth order with scale $\mu = (1.0,1.4,2.0.3.0)~{\rm GeV}^2$ are present in Table~\ref{t:xin_value}, where the results from other theoretical group are also given. Secondly, we give the $D\to\rho$ TFFs at large recoil region, {\it i.e.} $A_1(0), A_2(0)$ and $V(0)$ in Table~\ref{T:TFF0} and ratio $r_{2,V}$ in Fig.~\ref{F:TFFsratio}, which indicate our results are reasonable and consistent with many approaches within errors. Thirdly, we calculated the $|V_{cd}|$-independent total decay width, ratio for longitudinal/transverse and positive/negative helicity decay width results and present them in Table~\ref{T:Gamma}. A detail discussion in comparing with other predictions is made. Lastly, we give branching fraction of the two types $D\to\rho\ell^+\nu_\ell$ semileptonic decays in Table~\ref{T:BF}. In the near further, we hope more accuracy data will be reported and more theoretical results will be given to explain the gaps between different approaches.

\acknowledgments This work was supported in part by the National Natural Science Foundation of China under Grant No.12265009, No.12265010, No.12175025 and No.12147102, the Project of Guizhou Provincial Department of Science and Technology under Grant No.ZK[2021]024, No.ZK[2023]142, the Project of Guizhou Provincial Department of Education under Grant No.KY[2021]030, and by the Chongqing Graduate Research and Innovation Foundation under Grant No. ydstd1912. \\

\end{document}